# Performance analysis of frequency regulation services provided by aggregates of domestic thermostatically controlled loads

F. Conte[1,*], M. Crosa di Vergagni[1], S. Massucco[1], F. Silvestro[1], E. Ciapessoni[2], D. Cirio[2]

*1 Department of Electrical, Electronic, Telecommunication Engineering and Naval Architecture, University of Genova, Via all'opera Pia, 11a, 16145 Genova, Italy*

*2 Sviluppo dei Sistemi Energetici – SSE, RSE S.p.A – Ricerca sul Sistema Energetico, via R. Rubattino 54, 20134 Milano, Italy*

_______________________________________________________________________________

*Abstract*

   This paper proposes a control method for allowing aggregates of thermostatically controlled loads to provide synthetic inertia and primary frequency regulation services to the grid. The proposed control framework is fully distributed and basically consists in the modification of the thermostat logic as a function of the grid frequency. Three strategies are considered: in the first one, the load aggregate provides synthetic inertia by varying its active power demand proportionally to the frequency rate of change; in the second one, the load aggregate provides primary frequency regulation by varying its power demand proportionally to frequency; in the third one, the two services are combined. The performances of the proposed control solutions are analyzed in the forecasted scenario of the electric power system of Sardinia in 2030, characterized by a huge installation of wind and photovoltaic generation and no coil and combustible oil power plants. The considered load aggregate is composed by domestic refrigerators and water heaters. Results prove the effectiveness of the proposed approach and show that, in the particular case of refrigerators and water heaters, the contribution to the frequency regulation is more significant in the case of positive frequency variations. Finally, the correlation between the regulation performances and the level of penetration of the load aggregate with respect to the system total load is evaluated.

_______________________________________________________________________________

## 1. Introduction

   The progressive installation of renewable power plants has deeply modified the Italian and European electric power systems. As opposed to the traditional generating units, renewable generators are widely distributed throughout the electric network, hardly predictable and not dispatchable, and their inclusion in the provision of ancillary services to the grid is still an on-going process. In particular, considering frequency regulation services, the Italian electric grid is already characterized by reduced reserve margins, which depend on the number of synchronous generators on service [1]. In this context, the possibility of having new entities, other than the traditional generating units, enabled to provide frequency regulation services is becoming more urgent, if not compulsory [2].

   The possibility of including electric loads in the grid regulation and management activities is gaining attraction, since loads are already widely distributed. This approach is commonly referred to as Demand Side Response (DSR). Some examples of DSR strategies applied to residential loads can be found in [3]–[6]. To support frequency regulation, controlled loads would need to vary their power demand without compromising the final customer comfort. One of the most promising solutions is the active participation of thermostatically controlled loads (TCLs) to frequency regulation activities [7]–[9]. The scientific community has been very interested and active in this research topic. The authors in [10]–[19] consider an aggregate of TCLs in order to follow an arbitrary power profile, which can be thus any regulation signal, eventually defined by frequency measurements (*e.g.* the so called Area Control Error (ACE)). The control of TCLs is generally realized with three possible modalities: by a direct on\off switching, which is the case of [8], [10], [13], [14], [17]–[19], by varying the temperature set-point, which is the case of [7], [12], [16], and by an hybrid control that combines on\off switching and temperature set-point variation, which is the case of [11], [15].

The above-mentioned approaches are based on completely or partially centralized control strategies, which therefore assume the availability of real-time telecommunication infrastructures. In some cases, such as in [17]–[19], communication is not explicitly required but the local load management system (LLMS) should know information about the average dynamical behavior of the controlled TCLs aggregate, which should be updated. Moreover, in many cases, a certain level of computation capability is required to the LLMS.

To obtain an immediate implementation on domestic devices, which are already widely distributed and, at the moment, not connected to an efficient communication platform, a simple and fully decentralized control strategy is more advantageous. The European Network of Transmission System Operators for Electricity (ENTSO-E) proposes a decentralized approach in [20], defining the guidelines for future grid codes, suggesting simple rules for the frequency sensitivity of thermostatically controlled devices that could become mandatory in the near future. In particular, the ENTSO-E defines the guidelines for the control activation and deactivation with respect to a specific frequency dead-band, and it proposes a control strategy based on the variation of the temperature set-point of the thermostat proportionally to the frequency deviation from its nominal value. This approach has been studied in [21] and a similar method has been proposed and tested in the British electric grid for domestic refrigerators in [7].

In [22]–[24] the ENTSO-E procedure has been evaluated and tested using an accurate model of the Sardinian electric system. In particular, thermo-dynamic models of refrigerators and boilers have been developed and stochastic external signals that have an influence on their operation, such as the external air temperature and the hot water utilization profile, have been defined. In those works, the control strategy proposed by ENTSO-E has proved to enable the thermostatic loads to provide an effective support to primary frequency regulation (PFR), succeeding in providing for the decrease of the total regulating energy of the grid, due to the high level of renewable resources penetration. The control parameters have been optimized in [24], to obtain an efficient and robust contribution to PFR, without severe repercussions on secondary regulation. The potential degradation of the secondary regulation performance, which can be caused by the energy payback of the active loads that follows the power variation defined by the control strategy [19], [25], is a crucial point which can determine the effective applicability of the proposed control approaches. Specifically, a great variation of the load power demand, followed by a comparable recovery, can lead the frequency to overcome the control activation dead-band twice. The solution to this problem, as shown in [24], consists in limiting the frequency controller gain value. However, a great number of participating controlled devices could lead to a similar problem, even if the control gain is limited.

In this paper three alternative control logics, preliminary introduced in [26], are detailed and validated. They are not based on the variation of the thermostat temperature set-point, but on the forced activation or deactivation of the controllable devices, driven by the definition of a specific set of thresholds. The three strategies differ in the provided regulation services: in the first one, the TCLs aggregate power demand is enabled to vary proportionally to the Frequency Rate of Change (RoCoF), in order to provide synthetic inertia (SI); in the second one, the TCLs aggregate power demand is enabled to vary proportionally to frequency, providing, in this way, PFR; in the third one, the two services are combined. With respect to the methods introduced in [10]–[19], these control strategies are fully distributed, since they do not require any kind of communication among the controlled TCLs or with a central controller, but they only use local frequency measurements. Moreover, differently from the approach developed in [22]–[24], which, similarly, does not require communications, the TCLs aggregate effectively emulates inertia and the response of primary regulators of traditional generators.

In [26], a tuning of the control parameters is proposed, using a single scenario of the today Sardinian network. In the present paper, the method is deeply analyzed making use of a forecasted grid model of Sardinia for the year 2030, characterized by a huge installation of wind and photovoltaic generation and no coil and combustible oil power plants. Two load aggregates are considered, one composed by domestic refrigerators, that provides only PFR, and one composed by domestic water heaters, that provide both SI and PFR. Six scenarios with different network settings and ambient conditions, that determine different end-users requirements, are studied considering the occurrence of both over- and under-frequency events. Results prove the effectiveness of the proposed strategy and show that, in the considered case study, the contribution to frequency regulation is more significant in the case of positive frequency variations. Finally, a correlation between the regulation performances and the level of penetration of the load aggregate with respect to the system total load is shown.

It is worth remarking that, even if the control algorithm is distributed, the relevant contractual architecture may be not decentralized. Indeed, an aggregator could potentially measure frequency, establish the amount of primary reserve provided by the TCLs aggregate, and collect the eventual remuneration. However, the objective of the paper is to analyze the technical performances of the control strategies and how to remunerate the services is beyond the scope of the paper.

The reminder of the paper is organized as it follows: Section 2 introduces the three control strategies; Section 3 describes the TCLs models; Section 4 details the simulation scenarios; Section 5 reports the simulation results; Section 6 provides the conclusions of the paper.

## 2. Control strategies for frequency regulation services

The proposed control strategies, described in this section, are meant to manage the active power demand of an aggregate of $N$ TCLs with the objective of providing frequency regulation services. In particular, the considered regulation activities are:

1. the provision of SI, where the controllable loads aggregate would vary its active power demand proportionally to the RoCoF;
2. the provision of PFR, where the controllable loads aggregate would vary its active power demand proportionally to the frequency deviation from its nominal value;
3. a combination of the above-mentioned services of SI and PFR.

The proposed approaches do not change if the considered TCLs are cooling or heating units. We only assume that the $i$-th, ($i = 1,2, \ldots, N$) device is driven by a standard thermostat, which state $q_i$ can be equal to 0 (off) or to 1 (on). The thermostat controls a reference temperature indicated with $T_i$ [°C] and implements a standard logic with no state change within the dead-band $\left[T_i^d - \Delta, T_i^d + \Delta\right]$, where $T_i^d$ is the temperature set-point and $2\Delta$ is the dead-band extension. All the TCLs in the aggregate have the same nominal power $P^{\text{nom}}$ [W].

Moreover, all control schemes assume the use of local measurements of the grid frequency $f$ [Hz] and of RoCoF, hereafter indicated with $Df$ [Hz/s]. According to the ENTSO-E Demand Connection Code [20], frequency is assumed to be measured with a sampling time of 20 ms and filtered using a low-pass filter (with time constant $T = 100$ ms). The measured RoCoF, $Df^m$, is then computed as the difference between two consecutive frequency measurements, divided for the 20 ms time window.

It is worth remaking that, in this paper, PFR service is defined as usually done for traditional generators, *i.e.* as the provision of a power variation proportional to the frequency deviation from the nominal value, to be kept until secondary frequency controls return the system to nominal value (typically up to 15 minutes). Moreover, note that the response time obtained both for SI and PFR services is limited just by the above reported frequency measurement process. Therefore, the full response (meaning that the power variation is effectively proportional to the frequency derivative for SI and to frequency deviation for PFR with a 1% tolerance) is obtained in 520 ms (sum of the sampling time and $5 \cdot T$). This means that the provided services are compliant with the general definition of fast frequency response (FFR), provided for example in [27] as "power injected to (or absorbed from) the grid in response to changes in measured or observed frequency during the arresting phase of a frequency excursion event to improve the frequency nadir or initial rate-of-change of frequency". Always according to [27] the PFR service considered in this paper belong to the class of "sustained" FFR services since it is required to maintain the change in power injection until secondary frequency controls return the system to nominal frequency.

### A. Synthetic Inertia

To emulate inertia, an aggregate of loads should vary its total power demand proportionally to the RoCoF. In the case of TCLs, the thermostat states $q_i$ need to be modified according to the RoCoF value. The control scheme adopted in this work is reported in Figure 1.

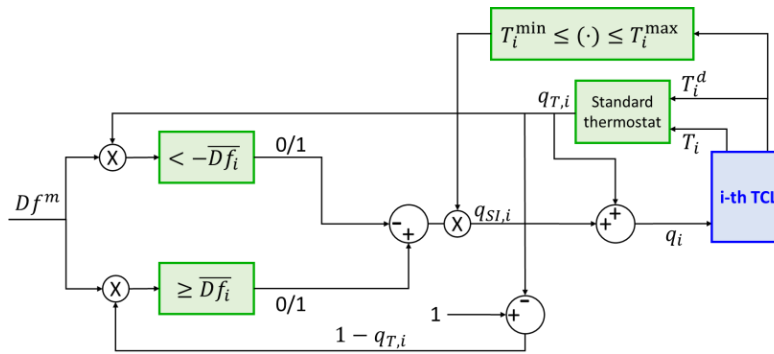

Figure 1 Block diagram of the inertia emulation method for the $i$-th TCL.

In Figure 1, we can observe that the thermostat state $q_i$ of the $i$-th TCL is given by

$$q_i = q_{T,i} + q_{SI,i} \qquad (1)$$

where $q_{T,i}$ is the state defined by the standard thermostat, and $q_{SI,i}$ is determined by the inertia emulation algorithm. Obviously, when $q_{SI,i} = 0$, the thermostat state is given by $q_{T,i}$. When the device is off ($q_{T,i} = 0$), the lower branch of the block diagram is activated since $1 - q_{T,i} = 1$. In this case, the measured RoCoF $Df^m$ is compared with the threshold $\overline{Df_i}$ [Hz/s]. If $Df^m \geq \overline{Df_i}$ (positive RoCoF event), $q_{SI,i}$ is set equal to 1, causing the activation of the device. When the device is on ($q_{T,i} = 1$), the higher branch of the block diagram is activated, and $Df^m$ is compared with $-\overline{Df_i}$. In this case, if $Df^m \leq -\overline{Df_i}$ (negative RoCoF event), $q_{IS,i}$ is set equal to -1, forcing the device deactivation.

In both cases, when $Df^m$ goes back inside the dead band $[-\overline{Df_i}, \overline{Df_i}]$, the thermostat state gets back to be defined by the standard logic. The upper-most block in Figure 1 stops the operation of the SI algorithm when the controlled temperature $T_i$ goes outside a certain *temperature security interval* $[T^{\min}, T^{\max}]$.

Figure 2 explains how an aggregate of TCLs, implementing the control scheme depicted in Figure 1, manages to provide SI. The thresholds $\overline{Df_i}$, $i = 1,2, ..., N$ are generated according to a uniform distribution between an activation threshold $Df_{act}$ [Hz/s] and a maximum value $\overline{Df}_{\max}$ [Hz/s], both fixed for all controlled devices, *i.e.* $\overline{Df_i} \sim U(Df_{act}, \overline{Df}_{\max})$ for all $i = 1,2, ..., N$. Given a certain working point, there would be $N^a$ on-devices and $N^d = N - N^a$ off-devices. Under the hypothesis that $N$ is sufficiently large, it is possible to assume that the separated sets of thresholds $\overline{Df_i}$, assigned to the on- and off-devices are again uniformly distributed in $[Df_{act}, \overline{Df}^{\max}]$. In Figure 2, to simplify the representation, and without loss of generality, the devices are ordered with the first $N^d$ off-devices and the remaining $N^a$ on-devices.

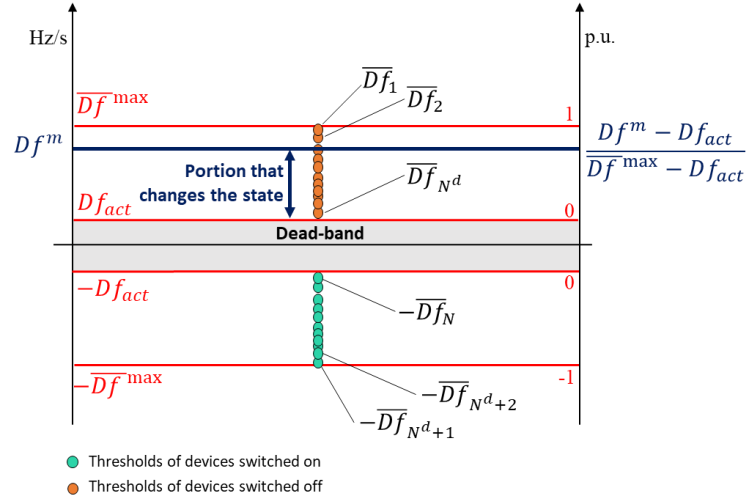

Figure 2 Synthetic inertia (SI) algorithm: power variation of the TCLs aggregate.

According to Figure 1, given a measured RoCoF value $Df^m \geq 0$, the $i$-th off-device would turn on if $Df^m \geq \overline{Df_i}$. Therefore, as shown in Figure 2, all the off-devices with a threshold value lower than $Df^m$ would change their state. Since the thresholds are uniformly distributed, quantities $q_{SI,i}$, that force the devices to switch on, are Bernoulli binary random variables, mutually independent and identically distributed, *i.e.* $q_{SI,i} \sim \mathcal{B}(r)$. Parameter $r$ is the probability of having $q_{SI,i} = 1$, which is equal to the probability of having $\overline{Df_i} \leq Df^m$. Therefore, since $\overline{Df_i} \sim U(Df_{act}, \overline{Df}_{\max})$, it follows that:

$$r(Df^m) := P\big(q_{SI,i} = 1 | Df^m\big) = P\big(\overline{Df_i} \leq Df^m\big) = \begin{cases} 0, & 0 \leq Df^m < Df_{act} \\ \dfrac{Df^m - Df_{act}}{\overline{Df}^{\max} - Df_{act}}, & Df^m \geq Df_{act} \end{cases}. \qquad (2)$$

Moreover, the number of devices that will turn on is equal to

$$n_{SI}^d = \sum_{i=1}^{N^d} q_{SI,i} \,. \tag{3}$$

It is well known that the sum of $n$ Bernoulli random variables is a binomial random variable $x$ with mean value $\mu = E(x) = nr$ and variance $\sigma^2 = nr(1-r)$. Moreover, with $n$ sufficiently large, it can be approximated by the mean value, *i.e.* $x \approx \mu$. Indeed, given a binomial distribution with mean $\mu$ and variance $\sigma^2 = nr(1-r)$, the coefficient of variation $c_v = \sigma/|\mu| = \sqrt{nr(1-r)}/(nr)$ goes to zero with $n$ going to infinity. In the specific case of $n_{SI}^d$, we have:

$$n_{SI}^d \approx N^d \cdot r(Df^m) = \begin{cases} 0, & 0 \le Df^m < Df_{act} \\ N^d \dfrac{Df^m - Df_{act}}{\overline{Df}^{\max} - Df_{act}}, & Df^m \ge Df_{act} \end{cases}, \tag{4}$$

from which it follows that the corresponding variation of power demand is

$$\Delta P_{SI}^d = P^{\mathrm{nom}} \cdot n_{SI}^d \approx \begin{cases} 0, & 0 \le Df^m < Df_{act} \\ M_{SI}^d(Df^m - Df_{act}), & Df^m \ge Df_{act} \end{cases}, \tag{5}$$

where

$$M_{SI}^d = \frac{N^d P^{\mathrm{nom}}}{\overline{Df}^{\max} - Df_{act}}, \tag{6}$$

is the resulting coefficient of proportionality.

Similarly, when $Df^m < 0$, it is possible to obtain

$$\Delta P_{SI}^a = P^{\mathrm{nom}} \cdot n_{SI}^a \approx \begin{cases} 0, & -Df_{act} < Df^m < 0 \\ -M_{SI}^a(|Df^m| - Df_{act}), & Df^m \le -Df_{act} \end{cases}, \tag{7}$$

where $M_{SI}^a = N^a P^{\mathrm{nom}}/(\overline{Df}^{\max} - Df_{act})$. To summarize, the power demand variation of the loads aggregate determined by the control action is:

$$\Delta P_{SI} \approx \begin{cases} -M_{SI}^a(|Df^m| - Df_{act}), & Df^m < -Df_{act} \\ 0, & -Df_{act} \le Df^m \le Df_{act} \\ M_{SI}^d(Df^m - Df_{act}), & Df_{act} \le Df^m \end{cases}, \tag{8}$$

which, as desired, is proportional to the measured RoCoF.

Coefficients $M_{SI}^d$ and $M_{SI}^a$ depend on the aggregate working point and the value of the difference $\overline{Df}^{\max} - Df_{act}$, which is a control parameter. This is clearer if $M_{SI}^d$ and $M_{SI}^a$ are expressed in per unit (p.u.) with respect to the nominal power of the TCLs aggregate $P^{\mathrm{nom}}N$:

$$\bar{M}_{SI}^d = \frac{N^d}{N} \cdot \frac{1}{\overline{Df}^{\max} - Df_{act}}, \qquad \bar{M}_{SI}^a = \frac{N^a}{N} \cdot \frac{1}{\overline{Df}^{\max} - Df_{act}} \,. \tag{9}$$

The activation threshold $Df_{act}$ has been introduced to offer the possibility of activating the provision of SI by the TCLs aggregate only for significant values of RoCoF. Indeed, if we set $Df_{act} = 0$, we obtain a behavior close to that of synchronous generators but many TCLs will switch from on to off, and vice versa, very often. For many classes of TCLs, this can be a problem since such a fast and frequent switching can damage some of their components. The idea of the present work is that SI is provided by TCLs only when large RoCoF values are detected. From the Transmission System Operator (TSO) point of view, this will result as an increase of the system inertia provided in "emergency conditions".

Moreover, it is worth remarking that to provide SI with the proposed strategy, any TCL is required to be able to switch form on to off and vice versa, within a sub-seconds dynamics. This is possible for some classes of TCLs, such as resistive electric water heaters, but not possible for other classes of TCLs, such as refrigerators, which need some time to activate the power consumption because of the use of compressors.

## B. Primary Frequency Regulation

In order to provide the PFR service, the power demand of the TCLs aggregate needs to vary proportionally to the frequency deviation from the nominal value $f^{\text{nom}}$ [Hz]. The control strategy proposed for the PFR it is similar to the SI control logic previously described, with the difference that the input signal is now the measured frequency deviation $\Delta f^m = f^m - f^{\text{nom}}$. The control scheme is shown in Figure 3: in this case, an on-device is deactivated if $\Delta f^m \leq -\overline{\Delta f_i}$ and an off-device is activated if $\Delta f^m \geq \overline{\Delta f_i}$.

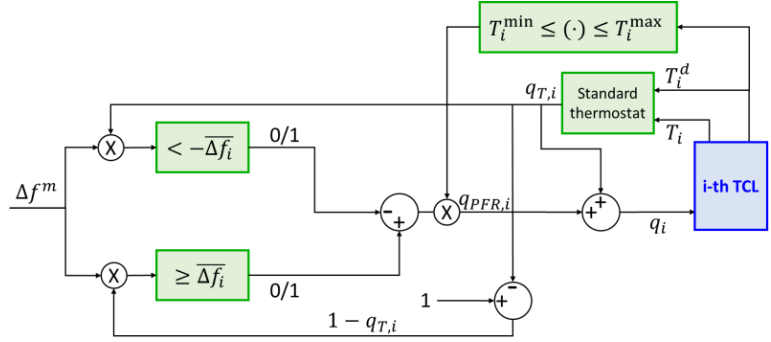

Figure 3 Block diagram of the PFR method for the $i$-th TCL.

Similarly to the SI case, each device has a frequency threshold $\overline{\Delta f_i}$, $i = 1,2,\dots,N$. These thresholds are generated according to a uniformly distributed between an activation threshold $\Delta f_{act}$ and the control parameter $\overline{\Delta f}^{\text{max}}$, *i.e.* $\overline{\Delta f_i} \sim U(\Delta f_{act}, \overline{\Delta f}^{\text{max}})$ for all $i = 1,2,\dots,N$. The control is inhibited if the temperature $T_i$ exceeds the security range $[T_i^{\text{min}}, T_i^{\text{max}}]$. Repeating the same mathematical steps described in Subsection 2.A, the load power variation results to be:

$$\Delta P_{PFR} \approx \begin{cases} -k_{PFR}^a(|\Delta f^m| - \Delta f_{act}), & \Delta f^m < -\Delta f_{act} \\ 0, & -\Delta f_{act} \leq \Delta f^m \leq \Delta f_{act}, \\ k_{PFR}^d(\Delta f^m - \Delta f_{act}), & \Delta f_{act} \leq \Delta f^m \end{cases} \tag{10}$$

where

$$k_{PFR}^d = \frac{N^d P^{\text{nom}}}{\overline{\Delta f}^{\text{max}} - \Delta f_{act}}, \qquad k_{PFR}^a = \frac{N^a P^{\text{nom}}}{\overline{\Delta f}^{\text{max}} - \Delta f_{act}}. \tag{11}$$

The corresponding p.u. values of gains $k_{PFR}^d$ and $k_{PFR}^a$ are

$$\bar{k}_{PFR}^d = \frac{N^d}{N} \cdot \frac{1}{\overline{\Delta f}^{\text{max}} - \Delta f_{act}}, \qquad \bar{k}_{PFR}^a = \frac{N^a}{N} \cdot \frac{1}{\overline{\Delta f}^{\text{max}} - \Delta f_{act}}. \tag{12}$$

As desired, the resulting load power variation is proportional to the frequency deviation. The equivalent control gains (12) depend on the aggregate working point and on the control parameter $\overline{\Delta f}^{\text{max}} - \Delta f_{act}$. The values of $\overline{\Delta f}^{\text{max}}$ and $\Delta f_{act}$ can be set in order to obtain a behavior similar to that of the traditional generators. Recall that, the control gain $k_g$ of a traditional generation unit with a nominal power of $P_g^{\text{nom}}$ is usually defined in order to get a specific value for the *droop* coefficient $b_p$ [p.u.] as $k_g = P_g^{\text{nom}}/(f^{\text{nom}} b_p)$. Therefore, the equivalent droop coefficients $b_p^d$ and $b_p^a$, defined with respect to the nominal power of the TCLs aggregate $N P^{\text{nom}}$ are:

$$b_p^d = \frac{N P^{\text{nom}}}{f^{\text{nom}} k_{PFR}^d} = \frac{\overline{\Delta f}^{\text{max}} - \Delta f_{act}}{f^{\text{nom}}} \cdot \frac{N}{N^d}, \qquad b_p^a = \frac{N P^{\text{nom}}}{f^{\text{nom}} k_{PFR}^a} = \frac{\overline{\Delta f}^{\text{max}} - \Delta f_{act}}{f^{\text{nom}}} \cdot \frac{N}{N^a}. \tag{13}$$

In (13), we observe that the equivalent droop coefficients are proportional to the control parameter $\overline{\Delta f}^{max} - \Delta f_{act}$. Note that the activation threshold $\Delta f_{act}$ introduces a frequency dead-band within which the TCLs aggregate does not provide PFR. The value of $\Delta f_{act}$ should be set according to the characteristic of the considered class of TCLs. Indeed, a value of $\Delta f_{act}$ close to zero, for example similar to that of traditional generators (0.01-0.02 Hz), will call the provision of PFR very often. As for SI, this means that many TCLs will often switch from on to off and vice-versa. How often depends on the dynamical characteristic of the grid (system inertia, types of loads, amount of renewable generations, etc.). For example, in the European Commission Regulation 2017/1458 [28], ±0.05 Hz is indicated as the standard frequency deviation interval for central Europe with the objective of violating such an interval for less than 15000 minutes per year (about 10 equivalent days). This means that, if in central Europe we set $\Delta f_{act}$ = 0.01-0.02 Hz, PFR will be required many times in one hour since we are within the standard frequency deviation interval. For some classes of TCLs, this can be a problem since a frequent on-off switching can damage electronical components. As for SI, the idea of the present work is that PFR is provided by TCLs only in "emergency conditions", when large frequency variations occur. In the mentioned case of central Europe, we can indicate as emergency threshold the value $\Delta f_{act}$ = 0.05 Hz.

### C. Combined SI-FPRF

The two control strategies previously described can be combined in order to obtain both contributions from the TCLs aggregate. The control scheme is the same of Figure 1 and Figure 3, but with the following signal as input:

$$\alpha(t)\Delta f^m + (1 - \alpha(t))Df^m \tag{14}$$

and, as thresholds,

$$\alpha(t)\overline{\Delta f_i} + (1 - \alpha(t))\overline{Df_i} \tag{15}$$

where $\alpha(t)$ is a time-varying coefficient. The initial value of $\alpha$ is zero; as a power imbalance causes a frequency variation, the controller starts operating having $Df^m$ as input signal, therefore implementing the SI control logic. As shown in Figure 4, after a certain pre-set amount of time, equal to $t_{switch}$ [s], $\alpha(t)$ starts increasing linearly with time and reaches the value of 1 after a transition interval of time $t_{ramp}$ [s].

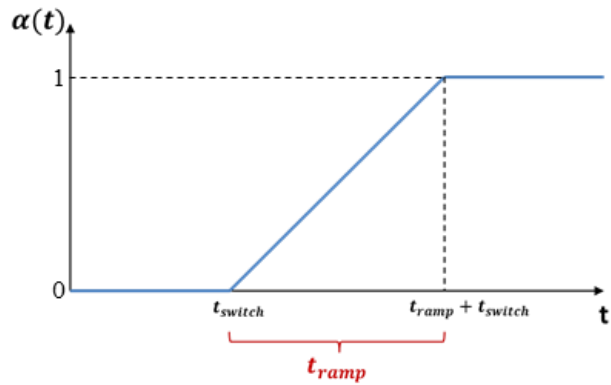

Figure 4 Time evolution of parameter $\alpha$.

When $\alpha(t) = 1$, the controller receives $\Delta f^m$ as input signal, and realizes the PFR control strategy. During transition, SI and PFR are operated simultaneously. In this case, two activation thresholds should be defined for RoCoF and frequency deviation. In particular, when the RoCoF exceeds the activation threshold $\pm Df_{act}$ the SI control is activated. If also the frequency deviation activation threshold $\pm\Delta f_{act}$ is violated, the ramp time evolution of $\alpha(t)$ is triggered.

## 3. Thermostatically Controlled Loads Models

In this paper, two type of TCLs are considered: domestic refrigerators and domestic electric water heaters. The adopted dynamical models are detailed in the following. This models are the same used in [22]–[24].

### A. Domestic refrigerators

A common domestic refrigeration system is compsed by a cooling compartment, a freezer compartment, and their respective contents. The thermal energy exchange scheme is shown Figure 5.

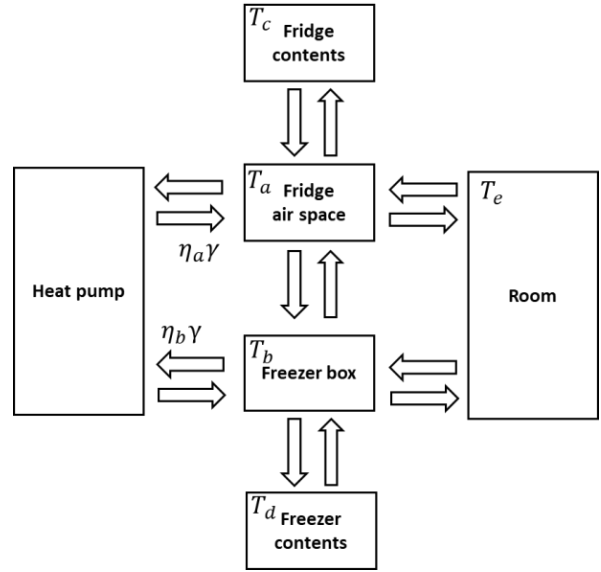

Figure 5 Block diagram of the domestic refrigerator thermal model.

The model is made of four controlled components (whose controlled temperature are denoted by $T_a, T_b, T_c$ and $T_d$ [°C]), the room external temperature $T_e$ [°C] and the heat pump. The latter extracts heat from both fridge and freezer compartments. There exist refrigeration systems with two independent heat pumps, but the ones with a single heat pump are the most common. To represent the thermal heat exchange performed by the heat pump, two positive variables with unitary sum $\eta_a$ and $\eta_b$ are introduced, which denote the separate share of heat absorbed by the fridge compartment and by the freezer compartment, respectively. Consequently, two equivalent coefficients of performace (COP) can be defined as $\gamma_a = \eta_a \gamma$ and $\gamma_b = \eta_b \gamma$, where $\gamma$ is the effective COP of the heat pump.

The dynamics of the temperatures is given by the following equations:

$$\dot{T}_a = -\frac{U_{a,b}A_{a,b}}{m_a S_a}(T_a - T_b) - \frac{U_{a,c}A_{a,c}}{m_a S_a}(T_a - T_c) - \frac{U_{a,e}A_{a,e}}{m_a S_a}(T_a - T_e) - \frac{1}{m_a S_a}\gamma_a q P^{\mathrm{nom}} \tag{16}$$

$$\dot{T}_b = -\frac{U_{a,b}A_{a,b}}{m_b S_b}(T_b - T_a) - \frac{U_{b,d}A_{b,d}}{m_b S_b}(T_b - T_d) - \frac{U_{b,e}A_{b,e}}{m_b S_b}(T_b - T_e) - \frac{1}{m_b S_b}\gamma_b q P^{\mathrm{nom}} \tag{17}$$

$$\dot{T}_c = -\frac{U_{a,c}A_{a,c}}{m_c S_c}(T_c - T_a) \tag{18}$$

$$\dot{T}_d = -\frac{U_{b,d}A_{b,d}}{m_d S_d}(T_d - T_b) \tag{19}$$

where: $m_x$ $(x = a, b, c, d)$ is the mass of the correspondig component [kg], $S_x$ is the specific thermal capacity [J kg$^{-1}$ °C$^{-1}$]; $U_{x,y}$ and $A_{x,y}$ are the thermal transimittance [W °C m$^{-2}$] and the area [m$^2$] of the heat exchange between the $x$ and the $y$ thermal components (the external ambient is considered to have infinite mass); $P^{\mathrm{nom}}$ is the heat pump nominal electric power [W]; $q$ is the thermostat state, which is driven by temperature $T_a$.

Model (16)-(19) is driven by one external input: the room external temperature $T_e$. This last depends on the external ambient temperature, which changes with seasons and daytime. Anyway, especially in winter, the internal house temperature is controlled by air heating/cooling systems. In this work, we use the approach introduced in [24] to define $T_e$, starting from the outdoor ambient temperature, taking into account the presence of air heating/cooling systems.

### B. Domestic electric water heater

A domestic water heating system, which will be referred to as boiler, is made of a storage space where the water is heated up by an electrical resistance exploiting the Joule effect, under the control of a thermostat. The heat exchange can be represented as shown in Figure 6.

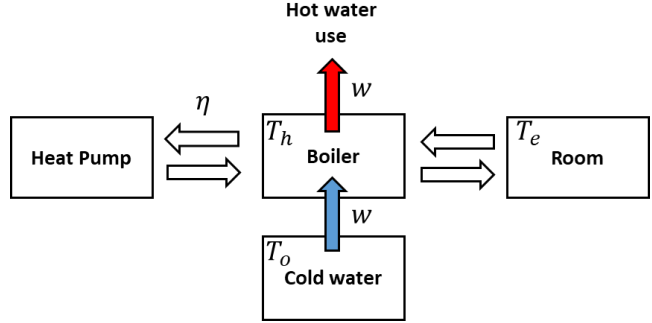

Figure 6 Block diagram of the domestic electric water heater (boiler) thermal model.

The internal temperature of the water within the boiler is assumed to be homogeneously equal to the variable $T_h$ [°C]. When the hot water is requested by the user, an equivalent cold water flow (denoted by $w(t)$ [m$^3$ s$^{-1}$]) enters the heating space. If the cold water temperature is equal to $T_o$ [°C], the thermal heat exchanged can be described by the following equation:

$$\dot{T}_h = -\frac{1}{R_{h,e} S_w V \rho}(T_h - T_e) - \frac{w(t)}{V}(T_h - T_o) + \frac{1}{S_w V \rho}\eta q P^{\text{nom}}, \qquad (20)$$

where: $T_e$ is the external room temperature [°C]; $R_{h,e}$ is the thermal resistance with respect to the thermal exchange with the outside [°C W$^{-1}$]; $T_o$ is the cold water temperature [°C]; $V$ is the volume of the boiler [m$^3$]; $S_w$ is the water specific thermal capacity [J kg$^{-1}$ °C$^{-1}$]; $\rho$ is the water density [kg m$^{-3}$]; $\eta$ and $P^{\text{nom}}$ [W] are the nominal efficiency and the nominal power, respectively; $q$ is the thermostat state.

Model (20) is driven by three external inputs: the room external temperature $T_e$, the cold water temperauture $T_o$, and the hot water use signal $w(t)$. In this work, the same approach of [24] is adopted to define these input variables, depending on the outdoor ambient temperature and taking into account the presence of air heating/cooling systems.

### C. TCLs aggregate model

The goal of the present study is the evaluation of the impact of a thermal loads flexible management on the frequency regulation activities, considering a regional or national electric network. The huge number of such loads makes the simulation of each single load not feasible. Thus, an equivalent aggregated model is required. In this paper, we use a Monte Carlo simulation approach, described in the following.

We consider two aggregates of TCLs, one composed by refrigerators and one composed by boilers. The aggregated nominal powers are $P_r^{\text{nom}}$ and $P_b^{\text{nom}}$ [W]. Each device belonging to the same aggregate is characterized by similar parameters, defined by a reference device model. For each class of TCLs, a number $m$ of sets of parameters with mean values equal to those of the reference model and a standard deviation $\sigma$ is generated. Each set identifies a load belonging to the considered class and it represents the behavior of a sub-aggregate of loads whose nominal power is equal to $P_x^{\text{nom}}/m$, with x = $r, b$. The sample number $m$ needs to be sufficiently low in order to allow the numerical simulation of all the loads aggregates (*e.g.* $\leq 1000$, depending on the available computational capabilities) and sufficiently high to represent the variability of the considered systems and the corresponding working conditions with adequate accuracy (*e.g.* $\geq 1000$).

The output of the numerical simulation of the set of aggregates are the thermostats state $q_{x,i}$, with x = $r, b$ and $i = 1,2, \ldots, m$. The total power absorbed by the set of loads of the class x is finally given by:

$$P_x = P_x^{\text{nom}} \sum_{i=1}^{m} \frac{q_{x,i}}{m}. \qquad (21)$$

### D. Adopted reference device models

In the following, the parameters of the reference device models adopted in this work for refrigerators and boilers are provided.

#### 1) Refrigerators

The refrigerator model considered as benchmark is the "Whirlpool WTE 31132 TS" with a capacity of 232/88 fridge/freezer. The parameters used in simulations have been chosen according to the technical data described in [29] and after a tuning performed through a series of simulations with the goal of reproducing specific temperature profiles similar to the experimental data reported in [30] and [31]. The resulting parameters are reported in Table 1. The values of the two equivalent COP $\gamma_a$ and $\gamma_b$, which refer to the thermal energy absorption capacity of the heat pump from the fridge and from the freezer, respectively, have been obtained considering an actual COP $\gamma = 1.2$ and considering the share absorbed from the fridge $\eta_1$ equal to the 38 % of that absorbed from the freezer.

Table 1 Parameters of the reference model for refrigerators.

| Component | Mass [kg] | Specific Heat Capacity [J/(kg °C)] |
|---|---|---|
| Refrigerator Air ($T_a$) | 10 | 2200 |
| Freezer Space ($T_b$) | 5 | 1000 |
| Refrigerator Content ($T_c$) | 10 | 4000 |
| Freezer Content ($T_d$) | 4 | 4000 |
| **Thermal Link** | **Area [m$^2$]** | **Transmittance [W °C m$^{-2}$]** |
| a-e | 2 | 0.5 |
| a-c | 1 | 12.5 |
| b-d | 0.26 | 2.5 |
| c-d | 0.4 | 12.5 |
| b-e | 0.97 | 0.15 |
| **Parameter** | **Symbol** | **Value** |
| Heat pump/fridge COP | $\gamma_a$ | 0.4560 |
| Heat pump/freezer COP | $\gamma_b$ | 0.7440 |
| Heat pump nominal power [W] | $P^{\text{nom}}$ | 100 |
| Thermostat dead-band [°C] | $2\Delta$ | 1 |

#### 2) Boilers

The water heater considered as benchmark is the Ariston TI-PLUS 100 V RTS/S, whose technical data is reported in Table 2.

Table 2 Parameters of the reference model for boilers.

| Description | Symbol | Value |
|---|---|---|
| Volume [l] | $V$ | 99 |
| Nominal power [W] | $P^{\text{nom}}$ | 1500 |
| Water specific heat capacity  [J kg$^{-1}$ °C$^{-1}$] | $S_w$ | 418.6 |
| Thermal resistance | $R_{h,e}$ | 0.777 |
| Heat-pump efficiency | $\eta$ | 1 |
| Thermostat dead-band | $2\Delta$ | 10 |

## 4. Case study

The test network considered in this work is the electrical network of Sardinia forecasted for the year 2030. The following Table 3 describes the technical characteristics of the network. The system components are grouped in: Sardinian generation units, Corse generation units, and HVDC links. Corse is included since the two islands are synchronously connected. The Sardinian generation units are classified in 3 groups: (G1) hydropower plants; (G2) gas turbines; (G3) biomass, solar thermal and equivalent plants; (G4) synchronous compensators; (G5) renewable energy sources; (G6) run-of-river hydro. Notice that there are no coil and combustible oil power plants, since they are not expected in the 2030 Sardinian electric network scenario.

The generation system in Corse is represented by three equivalent groups (hydropower, diesel ad gas turbine), which are considered to be connected to the Sardinian electric grid through the synchronous interconnection named SARCO. Finally, there are three HVDC links: two links with the Italian peninsula (SAPEI) and one link with both the Italian peninsula and Corse (SACOI).

Table 3 features, for each component, the following parameters: the active nominal power $P_{\text{nom},i}$ [MW] of generators or maximum imported power of the HVDC links; the minimum operating power $P_{\text{min},i}$ [MW], which is the minimum generated power of the generating units or the maximum exported power of the HVDC links; the generators start-up time $T_{s,i}$ [s], defined in [32] as a measure of the rotating inertia of the synchronous generators; the generators droop $d_i$ [%] (if the unit does not operate PFR the table reports "No"); the (half) frequency dead-band $\Delta f_i^{th}$ [mHz], which defines the frequency range within which primary regulation is disabled; the rate limit value $r_i^{\%}$ [%/min], defined as the maximal percentage variation (with respect to the nominal power) allowed to the units which are contributing to the secondary frequency control. Primary and secondary frequency control parameters are set based on the current Italian grid code [33].

Table 3 2030 Sardinian electrical network components parameters

| Unit | Type | Nominal power [MW] | Minimal power [MW] | Start-up time [s] | Primary control | | Secondary control |
|------|------|-------------------|-------------------|-------------------|------------------|--|-------------------|
| | | | | | Droop [%] | (Half) Dead-band [mHz] | Rate Limit [%/min] |
| | | $P_{\text{nom},i}$ | $P_{\text{min},i}$ | $T_{a,i}$ | $d_i$ | $\Delta f_i^{th}$ | $r_i^{\%}$ |
| **Sardinia Generation Units** | | | | | | | |
| Hydro (G1) | Hydro | 155 | 0 | 7.5 | 4% | 20 | 60% |
| Pumped Hydro (G1) | Hydro | 207 | -207 | 7.5 | No | No | No |
| UP2 (G2) | Gas Turbine | 100 | 25 | 15.3 | 5% | 10 | 8% |
| UP1 (G2) | CC-Gas Turbine | 80 | 24 | 9.4 | 5% | 10 | 8% |
| BioDisp (G3) | Biomass | 5 | 2 | 15.3 | 5% | 10 | 8% |
| Other thermal units (G3) | Thermal - Other | 127 | 51 | 14.9 | 5% | 10 | No |
| SARLUX (G3) | Equivalent | 550 | 165 | 9.4 | 5% | 10 | 8% |
| Codrrogianos 1 (G4) | Compensator | 250 | 0 | 3.5 | No | No | No |
| Codrrogianos 2 (G4) | Compensator | 250 | 0 | 3.5 | No | No | No |
| Photovoltaic (G5) | Photovoltaic | 2230 | 0 | - | No | No | No |
| Wind (G5) | Wind | 3250 | 0 | - | No | No | No |
| Bio Energetic (G5) | Bio Energetic | 50 | 0 | - | No | No | No |
| Run-of-river Hydro (G6) | Run-of-river Hydro | 17 | 0 | - | No | No | No |
| **Corse Generation Units** | | | | | | | |
| Diesel (G7) | Equivalent Diesel | 167.98 | 100 | 13.5 | 5% | 10 | 8% |
| Gas Turbine (G8) | Equivalent GT | 107.3 | 50 | 17.6 | 5% | 10 | 8% |
| Hydro (G9) | Equivalent Hydro | 125.8 | 0 | 8.1 | 5% | 10 | 8% |
| **HVDC Links** | | | | | | | |
| SAPEI 1 | HVDC | 500 | -500 | - | 5% | 20 | No |
| SAPEI 2 | HVDC | 500 | -500 | - | 5% | 20 | No |
| SACOI | HVDC | 300 | -300 | - | No | No | No |

## A. Electric Grid Model Implementation in MATLAB/Simulink™

An equivalent linearized model of the interconnected system Sardinia-Corse has been implemented in MATALAB/Simulink™. The model block diagram is reported in Figure 7.

Frequency dynamics is driven by the classical swing equation:

$$\Delta f(s) = \frac{f_{\text{nom}}}{T_s P_{\text{nom}}^N}\left(\sum_i P_{m,i}(s) - \sum_i P_{e,i}(s)\right), \tag{22}$$

where $P_{m,i}$ and $P_{e,i}$ [W] are the active power produced by the generating units and absorbed by the loads, respectively, $f_{\text{nom}} = 50$ Hz is the nominal frequency, $\Delta f$ [Hz] is the frequency deviation from its nominal power, $T_s$ [s] is the network start-up time, and $P_{\text{nom}}^N$ [W] is the grid nominal power. These two last quantities are computed as it follows, considering in the summations all synchronous generators:

$$P_{\text{nom}}^N = \sum_i P_{\text{nom},i}, \quad T_a = \sum_i T_{a,i}\frac{P_{\text{nom},i}}{P_{\text{nom}}^N}. \tag{23}$$

Generators speed control models has been implemented according the standard IEEE models HYGOV, for hydropower plants ((G1), (G9)), GAST for gas turbines ((G2), (G8)), and using equivalent models for Diesel and other generators ((G3), (G6), (G7)) by referring to DigSilent Power Factory™ [34] models.

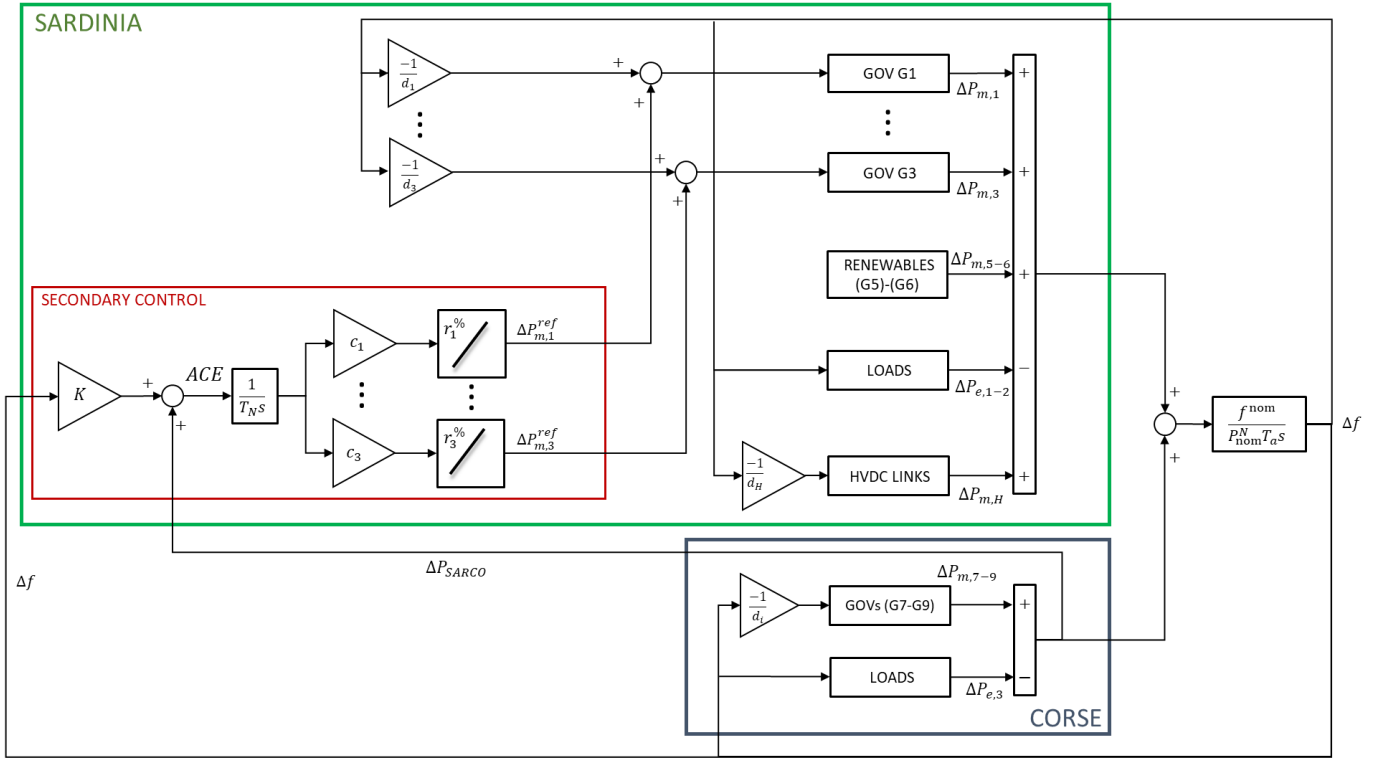

Figure 7 Network model block diagram.

PFR has been implemented according to the Italian grid code rules [33] for the conventional power plants using the relation:

$$\overline{\Delta P}_{m,i} = \frac{P_{\text{nom},i}}{f_{\text{nom}} d_i} \Delta f, \tag{24}$$

where $\overline{\Delta P}_{m,i}$ [W] is the requested mechanical power variation.

Also secondary frequency regulation has been implemented according to the Italian grid code rules [33]. The block diagram is reported in Figure 7. The objective is to recover the power unbalances and keep constant the synchronous AC power exchange with Corse ($\Delta P_{SARCO}$). The integration time constant $T_N$ has been set to 110 s. The integral gain $K$ is computed as

$$K = \sum_i \frac{P_{\text{nom},i}}{f_{\text{nom}} d_i}, \tag{25}$$

where the summation is over all connected generation units operating PFR. This parameter can be interpreted as regulating energy $E_r$ [MW/Hz], since it is equal to the static gain of the network power-frequency transfer function, *i.e.* $K = E_r = \Delta P / \Delta f_\infty$, where $\Delta f_\infty$ is the steady-state frequency deviation theoretically reached after the transient due to the occurrence of a power unbalance $\Delta P$. Roughly speaking $E_r$ is a measure of the frequency recovery capability of the network in a given operating conditions. Finally, $c_i$ with $i = 1,2,3$ are the secondary frequency regulation participation factors.

Loads are modeled with time varying signals. The load in Sardinia is divided between *not controllable* and *controllable*. The controllable one represent the refrigerators and boilers aggregates, modeled as described in Section 3, with $m = 1000$ samples and $\sigma = 10$ %. Boilers are able to operate SI, PFR and combined SI-PFR, introduced in Section 2. Refrigerators operate only PFR since, as discussed in Section 2.A, they are not adequate to provide SI. Table 4 reports the values adopted for control parameters. Note that two different activation thresholds $\Delta f_{act}$ are adopted for PFR of boilers and refrigerators. For boilers, $\Delta f_{act} = 0.05$ Hz, according to the definition of standard frequency deviation interval in central Europe, as reported in [28]. For refrigerators, $\Delta f_{act} = 0.1$ Hz. This choice is due to the technical characteristics of refrigerators, which can be damaged by a frequent and fast switching from on to off and vice versa. The idea is therefore that refrigerators are called to contribute to PFR only when frequency deviation is particularly severe. In other words, we can state that two levels of "emergency condition" are defined. The first level is reached when the standard frequency deviation interval of ±0.05 Hz is violated, and boilers start proving PFR. The second level is reached when frequency deviation crosses the threshold of ±0.1 Hz, and refrigerators start providing PFR.

Table 4 TCLs frequency control services parameters.

| Parameter | Symbol | Value | Algorithm |
|---|---|---|---|
| Max. RoCoF threshold for boilers | $\overline{Df}^{max}$ | 0.8 Hz/s | SI and SI-PFR |
| Max. frequency deviation threshold for boilers | $\overline{\Delta f}^{max}$ | 0.8 Hz | PFR and SI-PFR |
| Max. frequency deviation threshold for refrigerators | $\overline{\Delta f}^{max}$ | 0.8 Hz | PFR |
| SI activation threshold for boilers | $Df_{act}$ | 0.05 Hz/s | SI and SI-PFR |
| PFR activation threshold for boilers | $\Delta f_{act}$ | 0.05 Hz | PFR |
| PFR activation threshold for refrigerators | $\Delta f_{act}$ | 0.1 Hz | PFR and SI-PFR |
| Sample time frequency measure | $T_c$ | 0.02 s | All |
| Transition time from SI to PFR | $t_{switch}$ | 1 s | SI-PFR |

The general power dependency on frequency is modeled according to the following equation:

$$P_{e,i} = P_{e,i}^{ref} \left( 1 + \frac{K_{pf}}{\tau_{pf}s + 1} \right) \Delta f \, , \tag{26}$$

where $\tau_{pf} = 5$ s for all implemented aggregates models, $K_{pf} = 1.5$ s for the not controllable loads aggregates, $K_{pf} = 0.35$ s for refrigerators, and $K_{pf} = 0$ s for boilers (these values have been set according to Table 7.1 in [35]).

The HVDC links are modelled through import/export (positive/ negative) power profiles. The SAPEI power generation variation for PFR control is suitably modelled based on simplified converter model.

The implemented model has been validated by comparing the frequency profiles obtained within several different simulation scenarios with the ones obtained using a high-detailed model implemented on the DigSilent Power Factory™ [34] platform.

## B. Scenarios

Six scenarios are considered, characterized by different external temperatures conditions and network setups:

- Scenario A: nighttime (h=3 am), summer, external ambient temperature 19.6 °C;
- Scenario B: daytime (h = 10 am) summer, external ambient temperature 28.2 °C;
- Scenario C: evening (h = 10 pm) summer, external ambient temperature 23.4°C;
- Scenario D: nighttime (h = 3 am) winter, external ambient temperature 1 °C;
- Scenario E: daytime (h = 10 am) winter, external ambient temperature 4°C;
- Scenario F: evening (h = 10 pm) winter, external ambient temperature 3.3 °C.

Table 5 describes the grid components operating points and the secondary frequency regulation asset (participation factors), while Table 6 shows the grid general parameters: the grid start-up time $T_s$, the grid nominal power $P_{nom}^N$, both as above defined; the total power produced by the generating groups (synchronous machines and renewable plants); the power exchanged through the HVDC interconnections, where positive values stand for import and negative values for export; the power exchanged through the SARCO link; the regulating energy $E_r$; the upward and downward regulation margins for the PFR; and the *coefficients of penetration* of the TCLs aggregates for the over- and under-frequency events, denoted as $c_p^o$ and $c_p^u$.

These two last parameters are defined as it follows:

$$c_p^o = (1 - \rho) \cdot \rho_{nom}, \quad c_p^u = \rho \cdot \rho_{nom} \, , \tag{27}$$

where $\rho$ is defined as *the ratio between the operating point of the controllable load aggregate and the total load operating point in Sardinia*, and $\rho_{nom}$ is defined as *the ratio between the controllable load aggregate nominal power and the total load operating point in Sardinia*. We remark here that the amount of SI and of primary frequency reserve that a TCLs aggregate is ready to provide are given by (9) and (12), respectively. In these relations, we observe that SI and PFR, provided in case of over-frequency events, measured by $\overline{M}_{SI}^d$ and $\bar{k}_{PFR}^d$, are proportional to $N^d/N$, *i.e.* the number of off-devices ($N^d$) over the total number $N$ of the devices in the aggregate; whereas, in the case of under-frequency events, $\overline{M}_{SI}^a$ and $\bar{k}_{PFR}^a$ are proportional to $N^a/N$, *i.e.* the number of on-

Table 5 Simulation scenarios: operating points and regulation set-ups.

| Unit | Scenario A | | Scenario B | | Scenario C | | Scenario D | | Scenario E | | Scenario F | |
|---|---|---|---|---|---|---|---|---|---|---|---|---|
| | Operating Point [MW] | Secondary Regulation Part. Fact. | Operating Point [MW] | Secondary Regulation Part. Fact. | Operating Point [MW] | Secondary Regulation Part. Fact. | Operating Point [MW] | Secondary Regulation Part. Fact. | Operating Point [MW] | Secondary Regulation Part. Fact. | Operating Point [MW] | Secondary Regulation Part. Fact. |
| **Sardinia Generation Units** | | | | | | | | | | | | |
| Hydro (G1) | 155 | 0% | NIS | 0% | 155 | 0% | NIS | 0% | NIS | 0% | NIS | 0% |
| Pumped Hydro (G1) | NIS* | 0% | 13 | 0% | 146 | 0% | -166 | 0% | NIS | 0% | 166 | 0% |
| UP2 (G2) | NIS | 0% | NIS | 0% | NIS | 0% | NIS | 0% | NIS | 0% | NIS | 0% |
| UP1 (G2) | 40 | 10% | 41 | 10% | 52 | 10% | NIS | 0% | NIS | 0% | 42 | 0% |
| BioDisp (G3) | NIS | 0% | NIS | 0% | NIS | 0% | NIS | 0% | NIS | 0% | NIS | 0% |
| Other thermal units (G3) | 30 | 17% | 29 | 17% | 79 | 17% | 37 | 19% | 126 | 19% | 127 | 19% |
| SARLUX (G3) | 470 | 73% | 460 | 73% | 466 | 73% | 486 | 81% | 497 | 81% | 550 | 81% |
| Codrrogianos 1 (G4) | 0 | - | 0 | - | 0 | - | 0 | - | 0 | - | 0 | - |
| Codrrogianos 2 (G4) | 0 | - | 0 | - | 0 | - | 0 | - | 0 | - | 0 | - |
| Photovoltaic (G5) | 183 | - | 842 | - | 196 | - | 40 | - | 352 | - | 9 | - |
| Wind (G5) | 307 | - | 521 | - | 1247 | - | 326 | - | 929 | - | 113 | - |
| Bio Energetic (G5) | 50 | - | 50 | - | 50 | - | 50 | - | 50 | - | 50 | - |
| Run-of-river Hydro (G6) | 6 | - | 6 | - | 6 | - | 9 | - | 16 | - | 16 | - |
| **Corse Generation Units** | | | | | | | | | | | | |
| Diesel (G7) | 118 | - | 118 | - | 118 | - | 118 | - | 118 | - | 118 | - |
| Gas Turbine (G8) | 75 | - | 75 | - | 75 | - | 75 | - | 75 | - | 75 | - |
| Hydro (G9) | 88 | - | 88 | - | 88 | - | 88 | - | 88 | - | 88 | - |
| **HVDC Links** | | | | | | | | | | | | |
| SAPEI 1 | -165 | - | -257.5 | - | -350 | - | -112.5 | - | -345 | - | 15 | - |
| SAPEI 2 | -165 | - | -257.5 | - | -350 | - | -112.5 | - | -345 | - | 15 | - |
| SACOI | -100 | - | -150 | - | -150 | - | -72 | - | -150 | - | 8 | - |
| **Loads** | | | | | | | | | | | | |
| Not cont. load in Sardinia | -791 | - | -1180 | - | -1446 | - | -459 | - | -1002 | - | -935 | - |
| Controllable load in Sardinia | -25 | - | -116 | - | -101 | - | -26 | - | -128 | - | -126 | - |
| Load in Corse | -281 | - | -281 | - | -281 | - | -281 | - | -281 | - | -281 | - |

*NIS = Not In Service

Table 6 Simulation scenarios: grid general parameters.

| Parameters | Scenario A | Scenario B | Scenario C | Scenario D | Scenario E | Scenario F |
|---|---|---|---|---|---|---|
| Grid start-up time ($T_s$) [s] | 8.76 | 8.73 | 8.63 | 8.70 | 8.85 | 8.73 |
| Grid nominal power ($P_{nom}^N$) [MW] | 1813 | 1865 | 2020 | 1785 | 1578 | 1865 |
| Generated power [MW] | 1522 | 2242 | 2678 | 1063 | 2251 | 1304 |
| HVDC exchanged power [MW] | -425 | -665 | -850 | -297 | -840 | 38 |
| Total load (pumped hydro included) [MW] | -1097 | -1577 | -1828 | -766 | -1411 | -1342 |
| SARCO exchanged power [MW] | 0 | 0 | 0 | 0 | 0 | 0 |
| Regulating energy ($E_r$) [MW/Hz] | 940 | 966 | 1044 | 831 | 831 | 967 |
| Upward secondary reserve [MW] | 120 | 129 | 173 | 64 | 53 | 88 |
| Downward secondary reserve [MW] | 321 | 321 | 475 | 321 | 332 | 353 |
| Coefficient of penetration for over-freq. $c_p^o$ [%] | 52.4 | 30.9 | 26.6 | 86.5 | 34.7 | 36.5 |
| Coefficient of penetration for under-freq. $c_p^u$ [%] | 1.6 | 3.1 | 1.9 | 4.6 | 4.3 | 5.0 |

devices ($N^a$) over the total number of the devices in the aggregate $N$. Therefore, as mentioned in Section 2, the availability to increase or decrease the aggregate power demand depends on the aggregate working point. This last mainly depends on the external ambient temperature both for boilers and refrigerators, and also on the hour of the day for boilers, because of the different level of demand of hot water $w(t)$ (that in this paper is defined using the approach of [24]). Based on these considerations, penetration coefficients $c_p^o$ and $c_p^u$ defined in (27) represent a measure, relative to the total load of a given scenario, of the availability of a TCLs aggregate to increase or decrease its power demand in response to a positive or negative frequency variation.

In all scenarios, the aggregate of refrigerators and boilers has a total nominal power of 441 MW, 85 MW for refrigerators and 356 MW for boilers. Such values have been based on [36]. Observing the data in Table 6, we remark that all coefficients of penetration for under-frequency events $c_p^u$ have low values: from 1.6% in Scenario A to 5.0% in Scenario F. Differently, the values for over-frequency events $c_p^o$ are more significant: from 26.6% in Scenario C to 86.5% in Scenario D. This means that, in general, the potential response capability of the TCLs aggregate is high for over-frequency events and low for under-frequency events. This is due to the fact that, generally, the number of on-boilers and refrigerators is significantly lower than the number of off-devices.

*C. Simulations*

For each of the six scenarios, two "step" events are simulated, one causing frequency increase, one causing frequency decrease. Table 7 lists the performed simulations. All simulations last 30 minutes and the event occurs after one minute.

Table 7 Simulations list.

| Scenario | Event description | Amount [MW] | Event type |
|---|---|---|---|
| A | Loss of the SACOI link | 100 | Loss of exported power (over-frequency) |
| A | Wind disconnection | 100 | Loss of generation (under-frequency) |
| B | Loss of the SACOI link | 150 | Loss of exported power (over-frequency) |
| B | Wind disconnection | 100 | Loss of generation (under-frequency) |
| C | Loss of the SACOI link | 150 | Loss of exported power (over-frequency) |
| C | Wind disconnection | 100 | Loss of generation (under-frequency) |
| D | Loss of the SACOI link | 110 | Loss of exported power (over-frequency) |
| D | Bio-energy disconnection | 50 | Loss of generation (under-frequency) |
| E | Loss of the SACOI link | 150 | Loss of exported power (over-frequency) |
| E | Bio-energy disconnection | 50 | Loss of generation (under-frequency) |
| F | Load disconnection | 150 | Loss of demand (over-frequency) |
| F | Bio-energy disconnection | 50 | Loss of generation (under-frequency) |

## 5.   Results

Figure 8 - Figure 11 show the frequency profiles obtained during the first 25 seconds after the frequency events, in two example scenarios: Scenario A and Scenario B. In particular, the figures compare the frequency time evolution occurring without the contribution of the TCLs aggregate, and with the provision of the frequency control services introduced in Section 2: SI operated by boilers, FPRF operated by boilers and refrigerators and combined SI-PFR (with SI provided only by boilers). In Table 6, we can observe that Scenario A and Scenario B are characterized by a similar network start-up time $T_a$ and regulating energy $E_r$. Therefore, in the two scenarios, the grid has the same capability of response to frequency deviations in the case of no service from the TCLs aggregate. On the contrary, the coefficients of penetration of the TCLs aggregate are significantly different. Indeed, $c_p^o = 52.4\%$ in Scenario A and $c_p^o = 30.9\%$ in Scenario B, whereas $c_p^u = 1.6\%$ in Scenario A and $c_p^u = 3.1\%$ in Scenario B. This means that in Scenario A the TCLs aggregate has an higher capability of response to over-frequency events and a lower capability for under-frequency events, with respect to Scenario B.

Focusing first to the over-frequency events (Figure 8 and Figure 9), we can observe that the contribution of refrigerators and boilers allows the reduction of the maximal frequency deviation. The highest reduction is obtained with the SI-PFR control logic. In particular, frequency is kept lower than 50.2 Hz (around 50.15 Hz) in Scenario A and 50.3 Hz (around 50.23 Hz) in Scenario B. Notice that, in the no TCLs control case, in Scenario A, frequency reaches a value higher than 50.25 Hz, whereas, in Scenario B, it reaches a valuer higher than 50.35 Hz. The reduction of the maximal frequency deviation is obtained using only PFR is similar, even if slightly lower. Both with PFR and SI-PFR the quasi-steady-state frequency value, reached at the end of 25 seconds after the event, is reduced. Therefore, the contribution of TCLs implementing PFR and SI-PFR appears to effectively coincide with the one of a traditional generator providing droop PFR.

Both in Scenario A and Scenario B, the use of SI allows the reduction of the initial slope of frequency. This occurs in the only SI case, where the quasi-steady-state value is the same obtained in the no TCLs control case, and in the SI-PFR case, that thus successfully combines the provision of SI and PFR.

As expected, the effect of the TCLs contribution to the frequency regulation is less significant in the under-frequency cases (Figure 10 and Figure 11). Indeed, in Scenario A the difference among the four frequency profiles is hard to be observed, even if a slight reduction of the maximal frequency deviation occurs with all the three control strategies. In Scenario B, the effect of the TCLs response can be appreciated, even if less significant with respect to the over-frequency case.

Figure 12 - Figure 15 show how the different control solutions modify the power demand of the total TCLs aggregate in the two example scenarios Scenario A and Scenario B. In the over-frequency cases, SI gives a fast and significant contribution during the first 3 seconds, with a positive variation of power demand up to about 55 MW in Scenario A and to about 70 MW in Scenario B, and recovers the base load profile in less than 5 seconds. In the under-frequency cases, the contribution of SI is less significant, with a negative variation of power demand up to about 6 MW in Scenario A and to about 15 MW in Scenario B. Moreover, the base load profile is recovered in 10 seconds. PFR provides a slower load augmentation during the first seconds, but then this variation is maintained, as typically required by primary regulation. SI-PFR combines the two contributions. Also for PFR and SI-PFR, the power demand variation in the over-frequency cases is more significant.

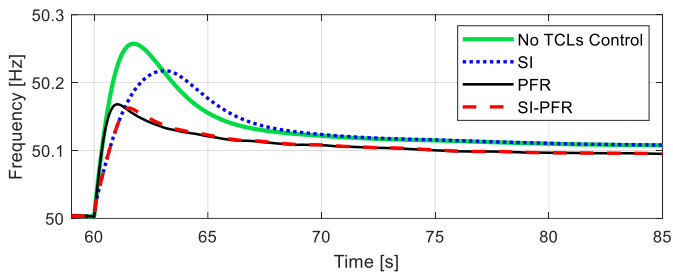

Figure 8 Scenario A: frequency during the first 25 seconds after the loss of the SACOI link (export of 100 MW).

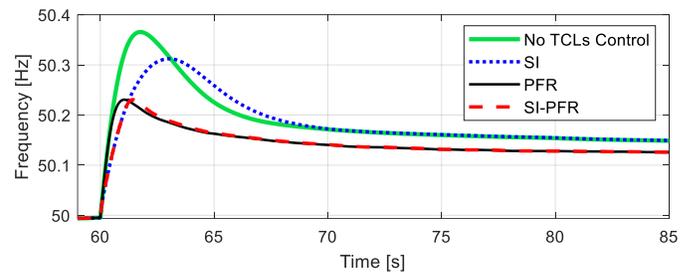

Figure 9 Scenario B: frequency during the first 25 seconds after the loss the SACOI link (export of 150 MW).

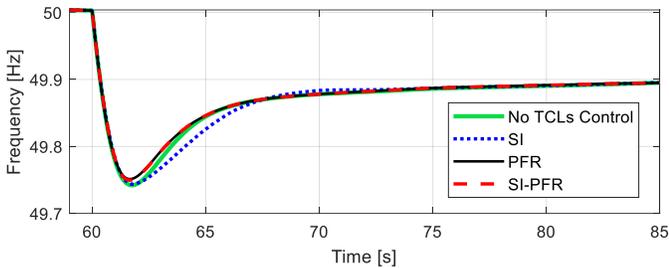

Figure 10 Scenario A: frequency during the first 25 seconds after the loss of wind generation (100 MW).

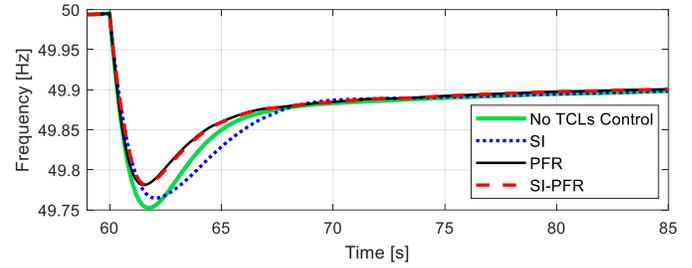

Figure 11 Scenario B: frequency during the first 25 seconds after the loss of wind generation (100 MW).

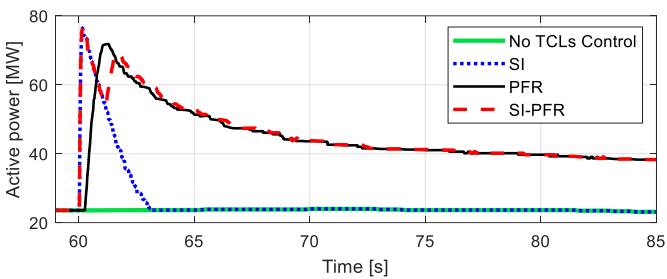

Figure 12 Scenario A: power demanded by the refrigerators-boilers aggregate during the first 25 seconds after the loss of the SACOI link (export of 100 MW).

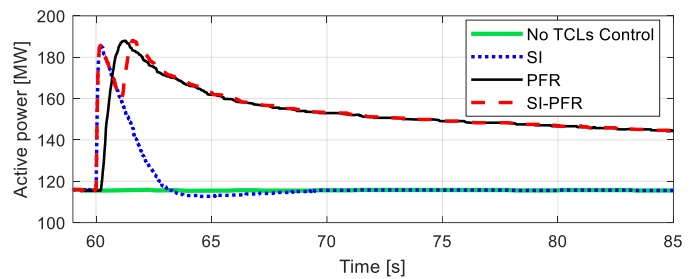

Figure 13 Scenario B: power demanded by the refrigerators-boilers aggregate during the first 25 seconds after the loss the SACOI link (export of 150 MW).

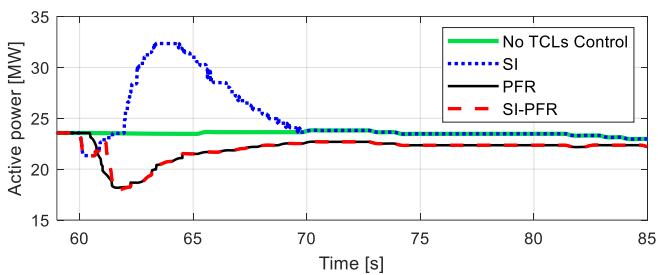

Figure 14 Scenario A: power demanded by the refrigerators-boilers aggregate during the first 25 seconds after the loss of wind generation (100 MW).

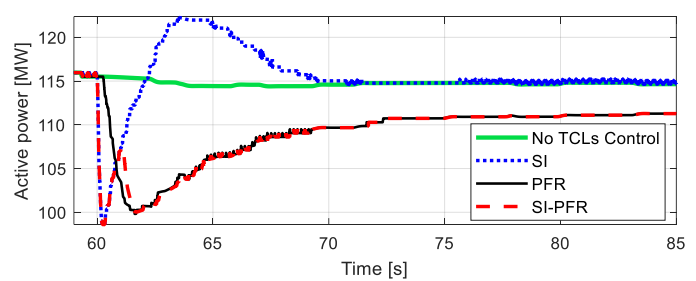

Figure 15 Scenario B: power demanded by the refrigerators-boilers aggregate during the first 25 seconds after the loss of wind generation (100 MW).

Notice that an immediate load payback occurs only in the SI case, after the under-frequency event (Figure 14 and Figure 15), without provoking significant variations on frequency. Moreover, this does not occurs in the PFR and SI-PFR cases. However, it is necessary to verify if, when the PFR service provision concludes, the energy recovery operated by TCLs can provoke further undesired frequency variations or underdamped oscillations (such as the *cold-load pickup oscillations* studied in [12]). To this aim, frequency and power demand are simulated within a 30 minutes time interval. Figure 16 and Figure 17 show frequency and TCLs power demand profiles within the entire 30 minutes of simulation in Scenario B. It results that after the initial transient, above analyzed, frequency recovers the nominal value in about 15 minutes thanks to secondary control with no significant differences among the TCLs no control case and the SI, PFR, and SI-PFR cases. This means that the load recovery does not compromise the secondary frequency regulation performance. At the end of the 30 minutes, both refrigerators and boilers are close to reach their base case power demand profiles. It is worth remarking that the same considerations are valid also for the other five scenarios.

Figure 16 and Figure 17 also allow the analysis the different contributions of boilers and refrigerators. In particular, in Figure 16, we observe that the variation from the base case power demand profile of the refrigerators aggregate ends around minute 2.5, when frequency deviation re-enters under the activation threshold $\Delta f_{act} = 0.1$ Hz (cyan dashed line). Refrigerators aggregate contribution to PFR is maximal during the initial transient with an augmentation of about 15 MW. The contribution of boilers is more significant, with a maximum augmentation of about 55 MW, and lasts more time, ending around minute 7, when frequency deviation re-enters under the relevant activation threshold $\Delta f_{act} = 0.05$ Hz (magenta dashed line).

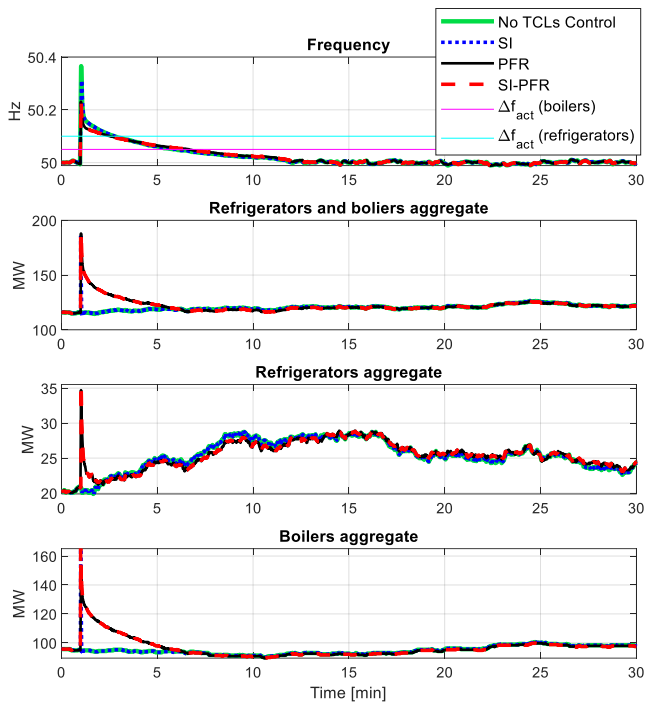

Figure 16 Scenario B: frequency and loads profiles after the loss of the SACOI link (export of 150 MW).

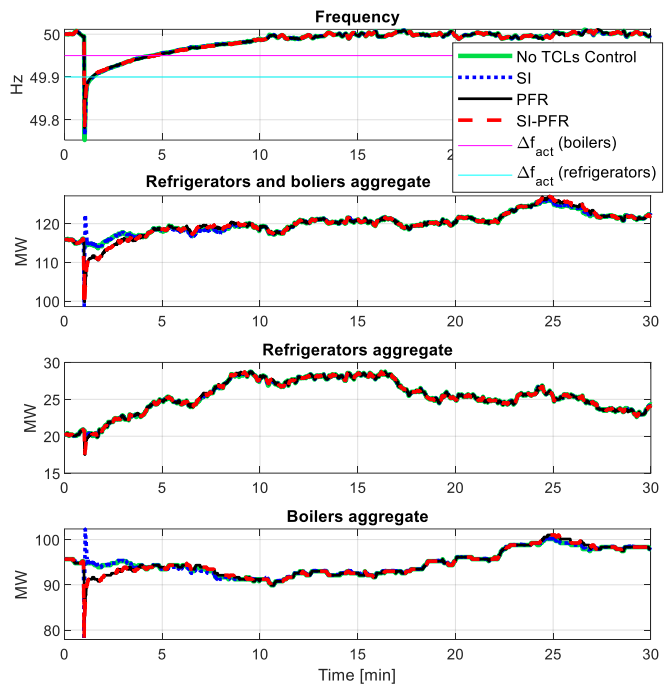

Figure 17 Scenario B: frequency and loads profiles after the loss of wind generation (100 MW).

Figure 18 and Figure 19 report the thermal dynamics of a set of sample boilers and refrigerators, respectively, in the case of Scenario B, after the loss of the export of 150 MW (over-frequency event), without the activation of the TCLs control and with SI-PFR. In Figure 18, we can observe the temperatures of the hot water of 50 boilers (randomly selected within the simulated $m = 1000$). Gray lines are the temperatures of boilers that do not modify their thermal dynamics because of the frequency variation, since their thresholds $\overline{\Delta f_i}$ and $\overline{Df_i}$ are higher than the values effectively reached by frequency and RoCoF. Colored lines are the temperatures that are deviated because of the activation of SI-PFR among the set of 50 boilers. For all of these boilers, we observe a slight variation between the case with no TCLs control (dashed lines) and with SI-PFR (dotted lines) that does not compromise the final user comfort. The bottom picture in Figure 18 reports the thermostat status of the set of boilers that contribute to SI-PFR. Here we can notice how these devices are temporarily activated within different time intervals, provoking, in some case (*e.g.* yellow line) a time shifting of the subsequent status changes. In Figure 19, we observe what happens in the case of a set of 50 sample refrigerators, that participate only to PFR. As for boilers, gray lines are the temperatures of refrigerators that are not called contribute to PFR, colored ones are those of refrigerators that are called to contribute to the service. Also in this case, temperature variations are limited and do not compromise the final user comfort. Similarly, in the bottom picture, we observe the temporary switching of the thermostats status and the consequent shifting of the following status changes. It is worth remarking that the thermal behavior of boilers and refrigerators in all the simulation settings is similar to the one above described for Scenario B after the loss of the SACOI link.

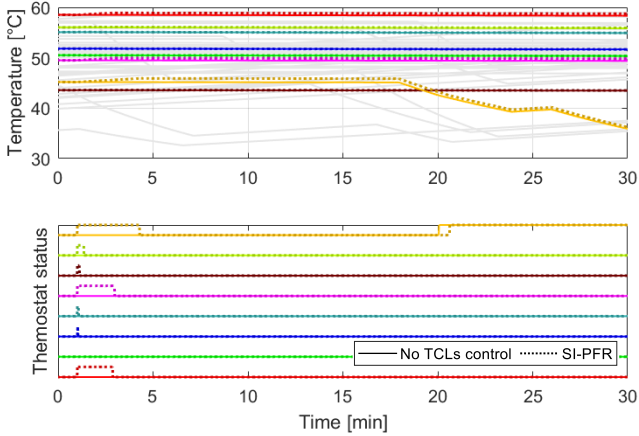

Figure 18 Scenario B: temperatures ($T_h$ in model (20)) and thermostat status of a set of sample boilers, after the loss of the SACOI link (export of 150 MW).

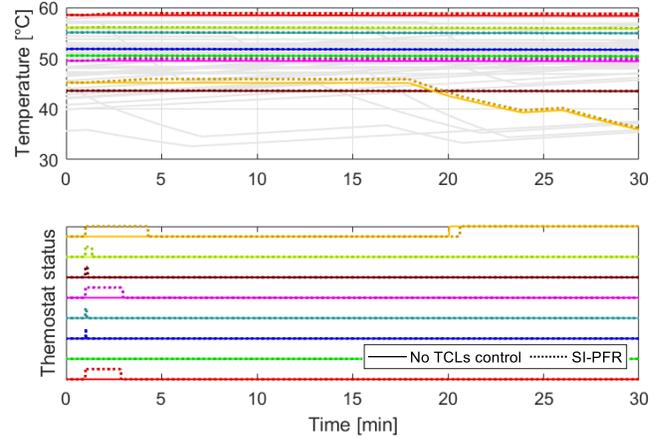

Figure 19 Scenario B: temperatures ($T_a$ in model (16)-(19)) and thermostat status of a set of sample refrigerators, after the loss of the SACOI link (export of 150 MW).

Figure 20 - Figure 27 graphically report the values obtained, for all the performed simulations, of the evaluation indices defined by UCTE in [37] (see also [23] for detailed definitions). For PFR [37] defines: $\Delta f_{max}$ [Hz], as the *maximal frequency deviation* (absolute value) reached after the event, and $\lambda_u$ [MJ], as the *network power frequency characteristic*, defined as the ratio between the power variation that causes the frequency event and the quasi-steady-state frequency deviation. For inertia, two indices are defined: $RoCoF_{100}$: frequency time derivative 100 ms after the perturbation, and $RoCoF_{500}$: frequency time derivative 500 ms after the perturbation.

Let us focus first on the results obtained for over-frequency events, reported in Figure 20-Figure 23. In Figure 20, we can observe that the contribution of the TCLs aggregate allows the reduction of the maximal frequency deviation $\Delta f_{max}$. Such a reduction goes from 0.04 Hz to 0.07 Hz with the SI service, from 0.06 Hz to 0.11 Hz with the PFR service, and from 0.07 Hz to 0.13 Hz with the combined SI-PFR service. In all scenarios, the largest reduction is obtained with SI-PFR, the lowest with SI.

The network power frequency characteristics $\lambda_u$ describes the grid capability to recover after a grid unbalance, since it is a measure of the equivalent primary regulating energy of the grid. As shown in Figure 21, the SI logic is not able to achieve any improvement in terms of $\lambda_u$, while the PFR and the SI-PFR strategies show the same results. In particular, the increment goes from 0.12 MJ in Scenario B to 0.16 MJ in Scenario D. The latter is, indeed, characterized by the highest controllable load penetration with respect to the total load in Sardinia (highest coefficient of penetration for over-frequency events $c_p^o = 86.5$ %, see Table 6).

Figure 22 and Figure 23 report the values of RoCoF measured after 100 ms and 500 ms from the frequency variation, respectively. Figure 22 clearly shows that the SI and SI-PFR control strategies reach the same significant improvements in terms of RoCoF reduction after the first 100ms. In particular, the best result is obtained in Scenario E, with a reduction of about 0.25 Hz/s. In Figure 23, we can observe that also the PFR method is able to grant a reduction in the RoCoF value after the first 500ms, even if the achieved result is not as significant as for the other two proposed control logics. In particular, the best result is again obtained in Scenario E, with a reduction of 0.09 Hz/s for the PFR logic and 0.14 Hz/s for both SI and SI-PFR.

The values of the evaluation indices reported in Figure 24-Figure 27 confirm the considerations given analyzing the frequency profiles for Scenario A and Scenario B (Figure 10 and Figure 11): in the under-frequency case, the capability of the aggregate of refrigerators and boilers to provide SI and PFR is limited. The maximal frequency deviation is reduced up to about 0.025 Hz in Scenario C with the SI-PFR, and of less significant values in the other scenarios. Similarly, the RoCoF after 100 ms and 500 ms is just slightly augmented with SI and SI-PFR. The result is not surprising since, as discussed above, in all scenarios, the coefficients of penetration for under-frequency events $c_p^u$ are lower than 5%.

Considering the entire set of results, we can conclude that the proposed strategies are well defined and they allow a TCLs aggregate to provide SI and PFR. However, the impact on the frequency dynamics depends on the load aggregate working point and on the level of penetration with respect to the total load. In the considered scenarios, the impact in the case of over-frequency events is significant, whereas the one in the case of under-frequency events is low.

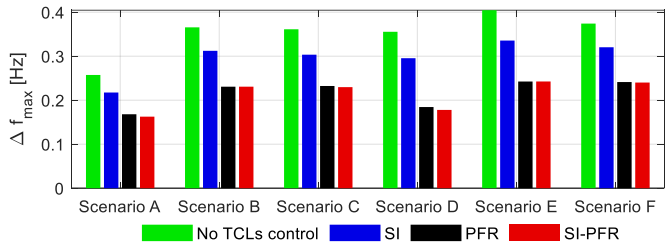

Figure 20 $\Delta f_{max}$ obtained with the different control strategies, for the over-frequency cases.

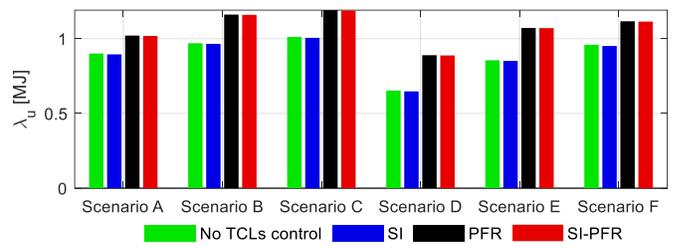

Figure 21 $\lambda_u$ obtained with the different control strategies, for the over-frequency cases.

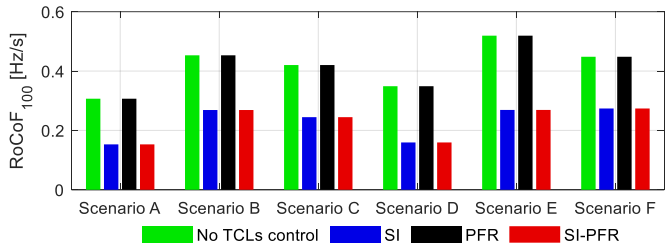

Figure 22 $RoCoF_{100}$ obtained with the different control strategies, for the over-frequency cases.

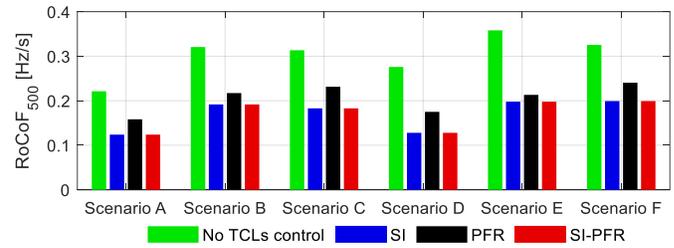

Figure 23 $RoCoF_{500}$ obtained with the different control strategies, for the over-frequency cases.

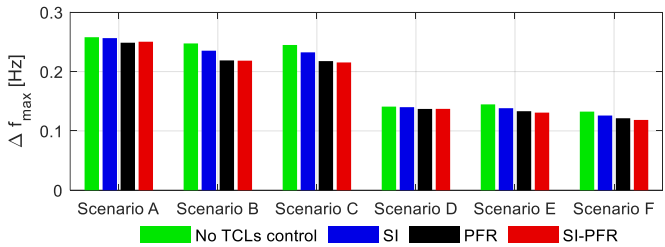

Figure 24 $\Delta f_{max}$ obtained with the different control strategies, for the under-frequency cases.

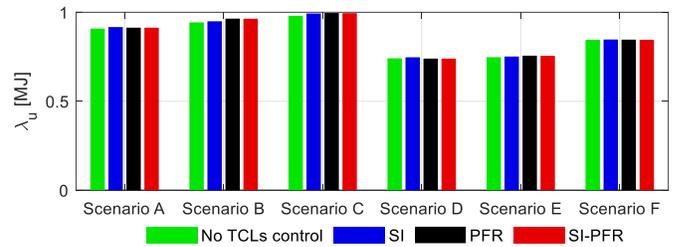

Figure 25 $\lambda_u$ obtained with the different control strategies, for the under-frequency cases.

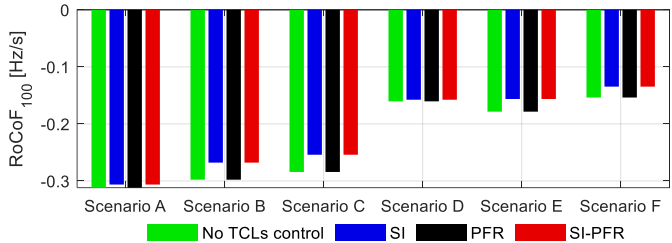

Figure 26 $RoCoF_{100}$ obtained with the different control strategies, for the under-frequency cases.

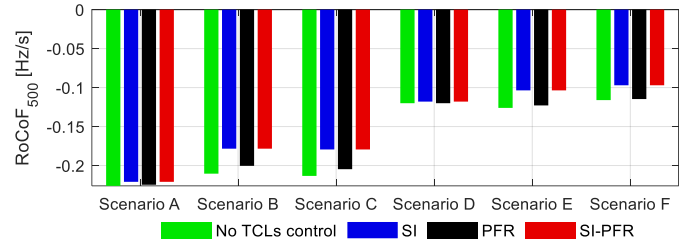

Figure 27 $RoCoF_{500}$ obtained with the different control strategies, for the under-frequency cases.

To further support these conclusions, an additive set of simulation implementing the SI-PRF strategy has been carried out by scaling up and down the nominal power of the refrigerators and boilers aggregate, obtaining five different couples of coefficients of penetrations for each scenario. To evaluate the correlation between $c_p^o$ and $c_p^u$ and the TCLs aggregate regulation performance, the following gains of performance are defined: $k_{\Delta f_{max}}$ [%]: the maximum frequency deviation $\Delta f_{max}$ percentage reduction, $k_{\lambda_u}$ [%]: the network power frequency characteristic $\lambda_u$ percentage increase; and $k_{RoCoF_{100}}$: the $RoCoF_{100}$ percentage reduction (in absolute value), all with respect to the no TCLs control case. Figure 28 - Figure 30 show the relation between the above-defined gains of performance and the penetration coefficients, obtained for the SI-PFR control strategy.

As expected, the performance gains improve with greater values of the penetration coefficients. We can observe that, by increasing the coefficient of penetration, the impact of the TCLs contribution to the frequency regulation can became significant also in the under-frequency case. For example, in some scenarios, with a $c_p^u$ close to 20%, it is possible to obtain a reduction of $\Delta f_{max}$ up to 20%, an increase of $\lambda_u$ up to 8%, and a reduction (in absolute value) of $RoCoF_{100}$ up to 25%.

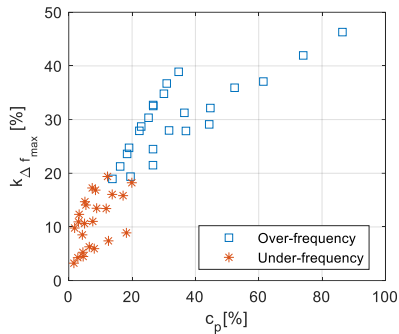

Figure 28 Relation between $c_p^{o/u}$ and $k_{\Delta f_{max}}$ for the SI-PFR strategy.

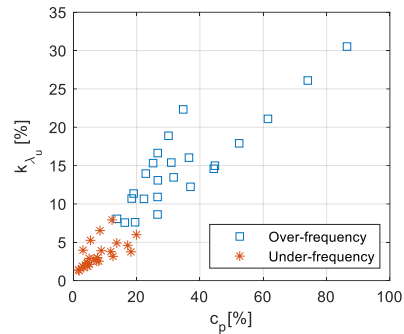

Figure 29 Relation between $c_p^{o/u}$ and $k_{\lambda_u}$ for the SI-PFR strategy.

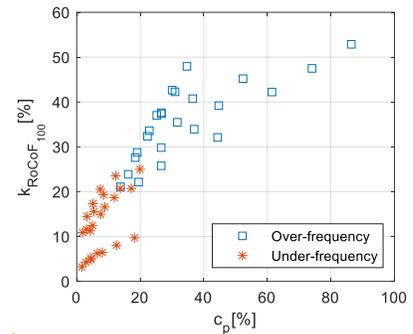

Figure 30 Relation between $c_p^{o/u}$ and $k_{RoCoF_{100}}$ for the SI-PFR strategy.

## 6. Conclusions

In this paper, a method for allowing aggregates of thermostatically controlled loads (TCLs) to provide synthetic inertia (SI) and primary frequency regulation (PFR) services has been proposed. The specific case of domestic refrigerators and electric water heaters has been considered and tested using a detailed model of the electric power system of an Italian island, forecasted for the year 2030. Six scenarios with different network configurations and end-users requirements have been simulated, considering the occurrence of over- and under- frequency events. The results show that the proposed strategies are well defined and they allow a TCLs aggregate to provide SI and PFR. However, an immediate and fast load payback occurs when only SI is implemented, but, if combined with PFR such a load payback is avoided. Moreover, the impact on the frequency dynamics depends on the load aggregate working point and on the level of penetration with respect to the total load. In the considered scenarios, the impact in the case of over-frequency events is significant, whereas the one in the case of under-frequency events is low.

**Acknowledgments**

This work has been financed by the Research Fund for the Italian Electrical System in compliance with the Decree of Minister of Economic Development April 16, 2018.